# Ballistic anisotropic magnetoresistance

J. Velev,[1] R. F. Sabirianov,[2] S. S. Jaswal,[1] and E. Y. Tsymbal[1*]

[1]*Department of Physics and Astronomy, University of Nebraska, Lincoln, Nebraska 68588-0111*
[2]*Department of Physics, University of Nebraska, Omaha, Nebraska 68182-0266*

Electronic transport in ferromagnetic ballistic conductors is predicted to exhibit ballistic anisotropic magnetoresistance (BAMR) – a change in the ballistic conductance with the direction of magnetization. This phenomenon originates from the effect of the spin-orbit interaction on the electronic band structure which leads to a change in the number of bands crossing the Fermi energy when the magnetization direction changes. We illustrate the significance of this phenomenon by performing *ab-initio* calculations of the ballistic conductance in ferromagnetic Ni and Fe nanowires which display a sizable BAMR when the magnetization changes direction from parallel to perpendicular to the wire axis.

The resistivity of bulk ferromagnetic metals depends on the relative angle between the electric current and the magnetization direction. This phenomenon was discovered by Thomson in 1857 and was called Anisotropic Magnetoresistance (AMR).[1] The importance of this phenomenon was recognized more than a century later in the 1970s when AMR of a few percent at room temperature was found in a number of alloys based on iron, cobalt, and nickel which stimulated the development of AMR sensors for magnetic recording (for reviews on AMR see Refs. [2] and [3]).

Ferromagnetic metals exhibiting a normal AMR effect show maximum resistivity when the current is parallel to the magnetization direction, $\rho_\parallel$, and minimum resistivity when the current is perpendicular to the magnetization direction, $\rho_\perp$. The magnitude of AMR can be defined by

$$AMR = \frac{\rho_\parallel - \rho_\perp}{\rho_\perp}. \qquad (1)$$

At intermediate angles between the current and magnetization direction, $\theta$, the resistivity of an AMR material is given by

$$\rho(\theta) = \rho_\perp + (\rho_\parallel - \rho_\perp)\cos^2\theta. \qquad (2)$$

The origin of AMR stems from the anisotropy of scattering produced by the spin–orbit interaction.[4] The resistance is largely controlled by the rate of scattering of the current-carrying *s* electrons into localized *d* states which in ferromagnetic materials, such as Co and Ni, are present at the Fermi energy primarily in the minority-spin channel. The spin-orbit interaction mixes the majority and minority spin channels which allows majority spin-electrons to scatter into *d* states thereby increasing resistivity. The effect is anisotropic because the admixture of the majority-spin *d* states into the minority-spin *d* states depends on the magnetization direction. This leads to larger scattering for electrons traveling parallel to magnetization and, therefore, to larger resistivity $\rho_\parallel$ compared to $\rho_\perp$ (see Refs. [2] and [3] for more details).

The mechanism of electronic transport changes dramatically in constrained geometries of the nanometer scale when the dimensions are reduced to be less than the mean free path of electrons. In this case electronic transport becomes ballistic rather than diffusive which is typical for macroscopic samples.[5] When the constriction width becomes comparable to the Fermi wavelength the conductance is quantized in units $2e^2/h$ for non-magnetic materials and in units $e^2/h$ for magnetic materials in which the exchange interaction lifts the spin degeneracy. The latter was observed in Ni break junctions,[6] Ni nanowires electrodeposited into pores of membranes,[7] and electrodeposited Ni nanocontacts grown between pre-patterned electrodes.[8] Recent experiments performed on Ni ballistic nanocontacts found a change of sign in the magnetoresistance obtained for the field parallel and perpendicular to the current.[9,10] This behavior was interpreted and signature of AMR.

The origin of magnetoresistance anisotropy in the ballistic transport regime is very different compared to that in the diffusive transport regime because there is no electron scattering contributing to the conductance in the former. The ballistic conductance is given by $G = Ne^2/h$, where $N$ is the number of open conducting channels, i.e. the number of transverse modes at the Fermi energy. This quantity is affected by the spin-orbit interaction which is known to be much stronger in open and constrained geometries than in bulk materials. The effect is anisotropic because the orbital momentum is coupled to the spin causing its projection to be different depending on the magnetization direction. By changing the magnetization direction one can, therefore, change the number of bands crossing the Fermi energy and thereby affect the ballistic conductance. We designate this phenomenon as the *ballistic anisotropic magnetoresistance* (BAMR) effect.

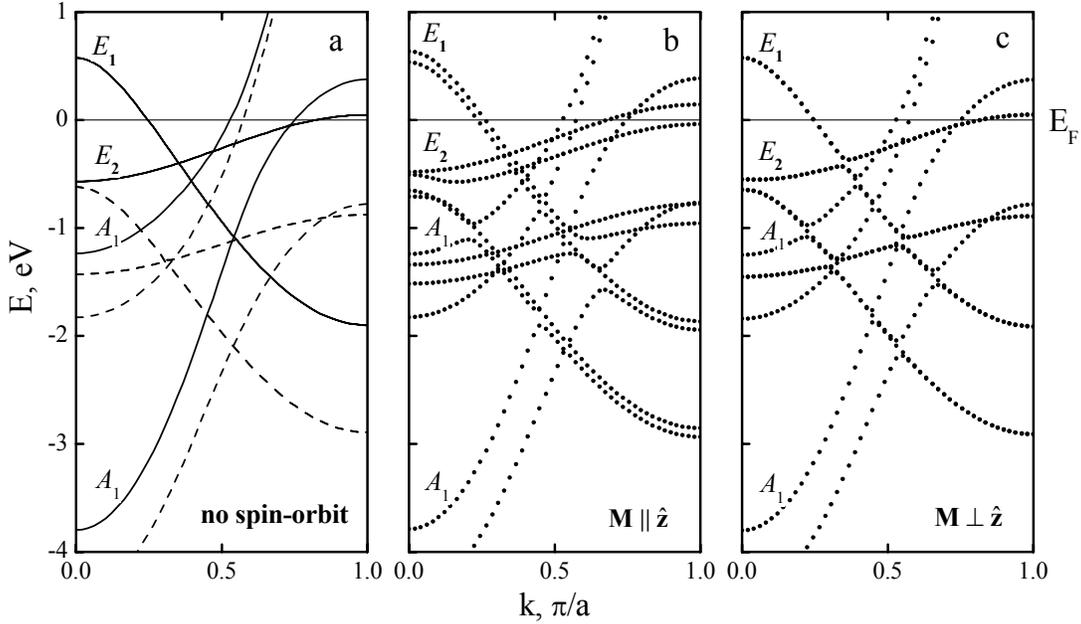

**Fig.1:** Calculated electronic structure of monoatomic Ni wire with equilibrium interatomic distance in the absence of spin-orbit interaction (a) and in the presence of spin-orbit interaction for magnetization lying along the wire axis $\mathbf{M} \parallel \hat{\mathbf{z}}$ (b) and perpendicular to the wire axis $\mathbf{M} \perp \hat{\mathbf{z}}$ (c). The solid and dashed lines in (a) show the minority - and majority-spin bands respectively. The labels stand for the irreducible representation of the group $C_{\infty\upsilon}$ and are displayed for minority-spin bands only.

In this Letter, we illustrate the significance of the BAMR effect by performing *ab-initio* calculations of the ballistic conductance of very thin ferromagnetic nanowires for magnetization parallel and perpendicular to the axis of the wire. We find that there is a sizable difference in the conductance for the two orientations of the magnetization giving rise to an appreciable BAMR. The BAMR effect stems from the spin-orbit interaction which lifts the degeneracy of the $d$ bands for the magnetization parallel but not perpendicular to the wire axis. This changes the number of conducting channels if the degenerate levels lie close to the Fermi energy. The BAMR is different from AMR observed in bulk materials because no electron scattering is responsible for it. We find that BAMR can be either positive or negative and predict a very different angular dependence compared to AMR.

We consider free standing, translationally invariant nanowires made of ferromagnetic fcc-nickel and bcc-iron, all having the tetragonal symmetry (except monatomic wires which have the axial symmetry). The nanowires are built along the [001] direction ($z$ axis) by periodic repetition of a supercell made up of two (001) planes (except for monatomic wires). We designate these wires as $m$-$n$ wires, where indexes $m$ and $n$ denote the number of atoms on each of the two topologically different layers in the (001) plane. The electronic structure of the ferromagnetic nanowires in the presence of the spin-orbit interaction is calculated using the pseudopotential plane-wave method[11] implemented within the Vienna *Ab-Initio* Simulation Package (VASP).[12] In order to use the advantage of the k-space representation within this method, we consider a periodic array of the wires separated by sufficiently large empty space to eliminate the coupling between the wires.

First, we discuss a monoatomic Ni wire representing a linear chain of Ni atoms with equilibrium lattice spacing of 2.17Å. Fig.1 shows the calculated electronic structure of this wire. In the absence of the spin-orbit coupling the electronic bands can be classified with no preferred magnetization orientation as minority- and majority-spin bands (the solid and dashed lines in Fig. 1a respectively). Due to the axial symmetry of the monoatomic wire these bands can be classified according to $C_{\infty\upsilon}$ group of the wavevector. This field splits the majority and minority $d$ bands into three subbands, two of which are doubly degenerate, labeled as $E_1$ and $E_2$ in Fig.1a, and one is non-degenerate, labeled as $A_1$ in Fig.1a. The labels stand for the irreducible representation of the group $C_{\infty\upsilon}$ given in Table 1. Note that the symmetry character of the bands is preserved throughout the whole Brillouin zone. The appearance of these states can be easily understood if we take into account that the crystal field of axial symmetry splits the $L = 2$ multiplet of the $d$ states into two doublets, $\{|+1\rangle \pm |-1\rangle\}/\sqrt{2}$ and $\{|+2\rangle \pm |-2\rangle\}/\sqrt{2}$, and a singlet $|0\rangle$.

The doublets have the $E_1$ and $E_2$ symmetry respectively and the singlet, belonging to the $A_1$ symmetry, is hybridized with the *s* state yielding the two bands of the $A_1$ symmetry.

**Table 1:** Symmetry character of the atomic orbitals in the group $C_{\infty v}$

| zx, yz | xy, $x^2-y^2$ | $3z^2 - r^2$ |
|--------|---------------|--------------|
| $E_1$  | $E_2$         | $A_1$        |

The spin-orbit interaction lifts the degeneracy of the doublets. However, the magnitude of the splitting is very different depending on the magnetization orientation with respect to the axis of the wire. In most general case the spin-orbit coupling has the form $H_{SO} = \lambda \mathbf{L} \cdot \mathbf{S}$, where the constant $\lambda$ is of the order of an meV. However, since the spin-orbit interaction is small compared to the spin splitting and the crystal field splitting of the bands, far from band crossings the $H_{SO}$ can be taken into account in the first order of perturbation theory. This gives rise to the respective effective operator in the form $H_{SO} = \pm\frac{1}{2}\lambda L_M$. Here $L_M$ is the component of the orbital momentum operator along the magnetization direction, and positive sign correspond to majority spins, whereas negative sign corresponds to minority spins. If the magnetization is parallel to the wire axis, $\mathbf{M} \parallel \hat{\mathbf{z}}$, the spin-orbit coupling is given by $H_{SO} = \pm\frac{1}{2}\lambda L_z$. Diagonalizing a 2x2 matrix of this operator within the $E_1$ doublet and the $E_2$ doublet independently, it is easy to find that the splitting of the $E_1$ doublet is $\lambda$, and the splitting of the $E_2$ doublet is $2\lambda$. These splittings are clearly seen in Fig.1b, demonstrating the strong influence of the spin-orbit interaction on the band structure of the wire when the magnetization lies along the wire axis.

This behavior changes dramatically when the magnetization is oriented perpendicular to the wire axis, say $\mathbf{M} \parallel \hat{\mathbf{x}}$. In this case the effective spin-orbit coupling is given by $H_{SO} = \pm\frac{1}{2}\lambda L_x$. It is easy to see that in this case the first order perturbation theory gives no contribution to the band splitting because the $L_x$ operator has zero matrix elements within the $E_1$ and $E_2$ doublets. The splitting occurs only in the second order and is, therefore, much smaller than for $\mathbf{M} \parallel \hat{\mathbf{z}}$. This is evident from Fig.1c which shows not much difference compared to Fig.1a where no spin-orbit interaction is taken into account. Only near the band crossing points in Fig.1c the splitting of the bands occurs, reflecting the effect of the spin-orbit coupling.

The influence of the magnetization orientation on the electronic band structure of the wire leads to BAMR. The ballistic conductance is controlled by the number of bands crossing the Fermi energy. As is seen from Fig. 1a, near the edge of the Brillouin zone the doubly degenerate $E_2$ band lies very close to the Fermi energy. The splitting of this band by the spin-orbit interaction, evident from Fig.1b, removes one band from the Fermi surface, reducing the number of conducting channels. Thus, one channel becomes closed for conduction as the magnetization orientation changes from $\mathbf{M} \perp \hat{\mathbf{z}}$ to $\mathbf{M} \parallel \hat{\mathbf{z}}$. This reduces the ballistic conductance by one quantum $e^2/h$ and hence results in positive BAMR. Using the BAMR ratio similar to Eq.(1) and keeping in mind that the total number of conducting channels for $\mathbf{M} \parallel \hat{\mathbf{z}}$ is equal 6, we find that the magnitude of BAMR in this case is 1/6 (or $\approx$17%).

Similar to the result for a monoatomic wire, our calculations performed for Ni wires of a larger cross sectional area show a tendency to have positive BAMR. For example, for a 5-4 Ni wire we find an increase in the conductance by one quantum when the magnetization orientation changes from the $\hat{\mathbf{z}}$ ([001]) to $\hat{\mathbf{x}}$ ([100]) direction, resulting in BAMR of 1/7 ($\approx$14%). This value, as well as the value of 17% obtained for a monoatomic Ni wire, is much larger than values of AMR in bulk materials being of the order of a few percent at room temperature. At the same time they are comparable to the magnetoresistance values observed in the experiments.[10] Also the predicted change in the conductance by one quantum due to the applied magnetic field is consistent with these experiments.

**Table 2:** Symmetry character of atomic orbitals in the group $C_{4v}$

| xy    | zx, yz, $L_x$, $L_y$ | $x^2-y^2$ | $3z^2 - r^2$ | $L_z$ |
|-------|----------------------|-----------|--------------|-------|
| $B_2$ | $E$                  | $B_1$     | $A_1$        | $A_2$ |

The mechanism of BAMR in nanowires of a larger cross section can be understood using arguments of the group theory. The group of the wave vector is now $C_{4v}$ whose irreducible representation are given in Table 2.[13] As is seen from this Table, there is a doubly degenerate *E* state corresponding to the *zx* and *yz* orbitals. In the first order of perturbation theory this state splits into singlets when the magnetization is parallel to the wire axis, but does not split if the magnetization is perpendicular to the axis. Indeed, if $\mathbf{M} \parallel \hat{\mathbf{z}}$ the spin-orbit coupling is given by $H_{SO} = \pm\frac{1}{2}\lambda L_z$. According to Table 2 the orbital momentum projection $L_z$ is transformed according to the $A_2$ representation. It has therefore non-zero matrix elements between the states of the $E$ doublet because the direct product $E \times E = A_1 + A_2 + B_1 + B_2$ contains the $A_2$ representation. This is opposite to the case of $\mathbf{M} \parallel \hat{\mathbf{x}}$ when the spin-orbit coupling is given by $H_{SO} = \pm\frac{1}{2}\lambda L_x$. In this case, as follows from Table 2, the orbital momentum projection $L_x$ transforms according to the $E$ representation which is not contained in the direct product $E \times E$. Therefore, there is no splitting of the doubly degenerate $E$ band when $\mathbf{M} \parallel \hat{\mathbf{x}}$.

Though the different band splittings for $\mathbf{M} \parallel \hat{\mathbf{z}}$ compared to $\mathbf{M} \parallel \hat{\mathbf{x}}$ are responsible for BAMR, they say nothing about sign of BAMR. The splittings can both enhance and reduce the ballistic conductance depending on whether bands are added to or removed from the Fermi surface. This fact is evident from our calculations of BAMR in Fe wires. Contrary to Ni wires, we find that Fe wires have a tendency for a negative BAMR. For both 4-1 and 9-4 Fe wires we find an opening of one additional conducting channel as the magnetization orientation changes from $\mathbf{M} \parallel \hat{\mathbf{x}}$ to $\mathbf{M} \parallel \hat{\mathbf{z}}$. The BAMR ratio in these cases is negative, equal to $-1/10$ ($-10\%$) and to $-1/16$ ($\approx -6\%$) for 4-1 and 9-4 Fe wires respectively. Note that a monoatomic Fe wire with the equilibrium lattice constant of 2.25 Å shows no BAMR.

The tendency of BAMR to have definite sign can be attributed to the position of flat degenerate bands with respect to the Fermi surface. We find that in case of Ni there are flat degenerate bands lying just above the Fermi energy and crossing the Fermi energy. Splitting of these bands by the spin-orbit interaction removes some bands from the Fermi surface and hence reduces the ballistic conductance leading to a positive BAMR. On the contrary, in Fe there are flat degenerate bands lying below the Fermi energy and not crossing the Fermi energy. The splitting of these bands by the spin-orbit interaction adds bands at the Fermi surface and hence enhances the ballistic conductance leading to a positive BAMR. This explains the tendency of opposite sign of BAMR in case of Ni and Fe. We note, however, that the complexity of the band structure and the appearance of band crossings near the Fermi energy might in certain cases change this tendency leading to a different sign of BAMR for wires of same material but different geometry.

The angular dependence of BAMR is very different compared to AMR. Since the ballistic conductance is an integer times $e^2/h$, the BAMR ratio is a step function of the magnetization angle. For example, in the case of monoatomic Ni wire we find that the conductance changes from $6e^2/h$ to $7\,e^2/h$ at an angle $\theta \approx 57°$ with respect to the wire axis. This behavior is very different from the angular dependence known for AMR in bulk samples given by Eq. (2). The predicted angular dependence of BAMR could be detected experimentally in ballistic magnetic nanocontacts displaying conductance quantization by measuring the conductance as a function of an angle of the applied magnetic field at saturation.

In conclusion, we have predicted the existence of ballistic anisotropic magnetoresistance (BAMR) – a change in the ballistic conductance with the direction of magnetization. This phenomenon originates from the effect of the spin-orbit interaction on the electronic band structure which leads to the change in the number of bands crossing the Fermi energy when the magnetization direction changes. This phenomenon is different from AMR observed in bulk materials because no electron scattering is responsible for it. Calculations performed for ferromagnetic Ni and Fe nanowires show a sizable change of the ballistic conductance when the magnetization changes the direction from parallel to perpendicular to the wire illustrating the BAMR effect. We find that BAMR can be either positive or negative and predict a very different angular dependence compared to AMR.

The authors thank Bernard Doudin for useful discussions. This work is supported by NSF (grants MRSEC: DMR-0213808 and DMR-0203359) and the Nebraska Research Initiative. Computations are performed using the Research Computing Facility at the University of Nebraska-Lincoln.

* Electronic address: tsymbal@unl.edu